\documentclass{article}     

\usepackage{graphicx}
\usepackage{subfigure}
\usepackage{textcomp}
\usepackage{multirow}
\usepackage{times}
\usepackage{amsmath}
\usepackage{courier}
\usepackage{type1cm}
\usepackage{flushend}
\usepackage{cite}
\usepackage{authblk}

\newcommand{\reged}{\textsuperscript{\textregistered\ }}

\newcommand{\SecureD}{\emph{SecureD}}

\date{}

\begin{document}

\title{\SecureD: A Secure Dual Core Embedded Processor}
\author[1]{Roshan G. Ragel}
\author[2]{Jude A. Ambrose}
\author[2]{Sri Parameswaran}

\affil[1]{Department of Computer Engineering, University of Peradeniya 20400 Sri Lanka}

\affil[2]{School of Computer Science and Engineering, University of New South Wales	NSW 2052 Sydney Australia}
\maketitle

\begin{abstract}
Security of  embedded  computing     systems is becoming of paramount
concern as these    devices  become more ubiquitous, contain personal
information and are increasingly used for     financial transactions.
Security attacks targeting  embedded systems illegally gain access to
the information in these devices or destroy information. The two most
common types of attacks embedded systems encounter are code-injection
and power analysis attacks. In the past, a number of countermeasures,
both hardware- and software-based, were proposed individually against
these two types of attacks. However, no single system      exists  to
counter both of these two prominent  attacks in a  processor    based
embedded system.  Therefore, this      paper, for the
first time, proposes a hardware/software based countermeasure against
both code-injection attacks and power analysis     based side-channel
attacks in a dual core embedded system. The proposed processor,
named \SecureD, has an area overhead of just 3.80\% and an    average
runtime   increase      of   20.0\% when compared to a standard dual
processing system. The overhead were measured using a set of industry
standard application benchmarks, with two encryption and five  other
programs.
\end{abstract}

\section*{Keywords}
Side Channel Attacks \and Code Injection Attacks \and Hardware/Software Co-design

\section{Introduction} \label{introduction}

Multiple processors have been used on a single chip as the increasing
functionality  of  embedded  systems  demands more processing  power.
Consumer
devices such as mobile phones already deploy dual processors in their
designs \cite{wolf05MultProc}. An increasing number  of such embedded
systems have to overcome security       threats. Potential targets of 
adversaries vary from low-end      systems such as wireless handsets,
networked sensors and smart cards,       to high-end systems such as
network routers, gateways, firewalls and servers.

Security threats in embedded systems could be classified by the means
used to launch attacks. Typical launch methods are: physical, logical
and side-channel. Physical attacks  refer  to  unauthorized  physical
access to the embedded system itself and are feasible   only when the
attacker has direct access to the system. Such attacks are identified
by  packaging  mechanisms,  such  as  tamper evident packing (make an
attempted attack apparent so that subsequent inspection will  show an
attack had been attempted) \cite{weingart00physical}. Logical attacks
exploit  weaknesses  in   logical   systems such as software   or   a
cryptographic  protocol  to  gain access to unauthorized information.
Logical  attacks are deployed easily  against  systems which are able
to download and  execute  software and have  vulnerabilities in their
design. Side-channel  attacks   are performed by observing properties
of the system (such as power consumption or electromagnetic emission)
while  the  system performs cryptographic operations. Unlike physical
attacks, logical and side-channel attacks  are  application  specific
and could be diagnosed and prevented by software and/or architectural
techniques.

Most  recent      logical attacks result in demolishing code
integrity of an application program \cite{milenkovic05hardware}. They
dynamically change instructions with the intention of gaining control
over the execution flow of a program.  Attacks  that  are involved in
violating software/code integrity are called  code-injection attacks.
Code-injection attacks often exploit common  implementation  mistakes
in  application programs,  which  are  referred to as        security
vulnerabilities.
The number of malicious attacks always increases  with  the amount of
software code \cite{cert07vulnerability}. The most
known side-channel attack is the power analysis attack
\cite{kocher99differential}, where   secret  keys used
in an encryption program were successfully predicted by observing the
power   dissipation      from a chip. Devices like \emph{smart cards}
\cite{biham99AESattack,chaumette}, \emph{PDAs} \cite{gebotys03ECCdummy}
and \emph{mobile phones} \cite{wolf05MultProc} have    microprocessor
chips built inside, performing secure transactions using secret keys.
Power dissipation/consumption  of  a   chip     is the most exploited
property used to predict  secret  keys  using    side channel attacks
\cite{koeune05ParallelComputing, zhou05side}.

A number of hardware- and software-based  techniques, such         as
ARM\reged \emph{TrustZone}    \cite{armtrustzone}    and \emph{ARMOR}
\cite{armor} respectively, were proposed in the past as        secure
infrastructures. However, they are not designed to detect/prevent the
most frequently  encountered  logical  and  side-channel  attacks  on
embedded systems: code-injection and   power  analysis  attacks.
There exist a  number     of individual detection
mechanisms  for  these  two  attacks  in  both   software         and 
architecture domains. However, to our knowledge, no single system was
proposed that is  capable of thwarting or warning against both  these
attacks.
Therefore, this paper, \emph{for the first time},   presents a secure
dual core  embedded  system, we call \SecureD\ (for {\bf Secure D}ual
core),   which  both detects
code-injection  attacks  and   protects      against            power
analysis attacks. \SecureD\  considers security as one of its  design
objectives   (as     opposed to security being an afterthought in the
design) \cite{kocher04security} and uses a hardware/software solution
with microarchitectural changes to the  processor design, which makes
the integration  of  security into the design and implementation much
faster and easier, while considerably reducing the overhead.

\subsection*{Paper Organization}
The  rest of the paper is organized as follows. Section~\ref{relwork}
summarizes   previously      proposed countermeasures  against   both
code-injection and power analysis attacks. The      microarchitecture
description  of  \SecureD\        is  provided in Section~\ref{arch}.
Section~\ref{designflow} illustrates   the design flow of our system.
The experimental setup is presented in Section~\ref{experSetup}   and
the  results  are  presented  in Section~\ref{results}. Finally,  the
paper is concluded in Section~\ref{conclusions}.

\section{Related Work} \label{relwork}
In this section, we present a range of prior detection and prevention
techniques, both hardware- and software-based            against both
code-injection and power analysis attacks. 

A wide range of techniques have been suggested in the past to detect/%
counter code-injection attacks. They  could  broadly  be  categorized
into  software based and hardware assisted techniques. Software-based
techniques use software tools and methods  to  overcome these attacks
without changing the microarchitecture of the processor.     Hardware
assisted                 techniques use additional hardware blocks or
microarchitectural  support to
detect code-injection attacks. Software-based techniques can     be
further categorized into two: static and dynamic.    Static 
techniques  try to  detect vulnerabilities at compile  time.   Wagner
et al.\ propose an automated static code analysis tool to detect code
that might  invite buffer overflow attacks \cite{wagner00first},  but
this produces a relatively large number   of false positives. Another
static technique uses a language that has only the safe constructs of
another language, for example, Cyclone  \cite{jim02cyclone}.   Dynamic
software based techniques avoid or considerably reduce code injection
attacks at runtime and they either use formal methods to  prove     a
program  behaves  as  expected or use software  constructs to monitor
proper  program     behaviour at runtime. \emph{Proof-Carrying  Code}
\cite{necula97proofcarrying} is  an  example   of    the  former and
\emph{Stack Guard} \cite{cowan98stackguard} the latter.

Hardware assisted techniques are mainly attack-specific. A number of
researchers \cite{lee03enlisting,mcgregor03processor,xu02architecture}
propose  architectural detection of buffer overflow attacks which are
a type of code-injection     attacks. Code-injection attacks can be
detected by monitoring code integrity of a
program  at    runtime.   Arora  et al.\ \cite{arora05secure} use  an
additional co-processor and hardware tables to perform  software/code
integrity  checks. This       system identifies program properties at
different levels of      granularity  and stores multiple control flow
levels   of    data    and   checksums to perform  software integrity
monitoring. This method produces code that  is not relocatable due to
the pre-generated    hardware tables. Ragel and Parameswaran in their
secure processor called \emph{IMPRES} \cite{ragel06impres}        use
similar  program properties to that of \cite{arora05secure}, but only
perform  check-summing at the basic block level and therefore  reduce
the complexity of their solution. However, \emph{IMPRES} only detects
code-injection  attacks and is designed for a single processor embedded
system~\cite{ragel2006architectural}.

Similar  to  detection   techniques   for   code-injection   attacks,
countermeasures against power analysis attacks can be classified into
software-based and  hardware-based. Masking and
current flattening are the  two major software-based countermeasures.
Table and data masking techniques \cite{coron00masking,gebotys06masking,Goubin99DPAmask,messerges02SMART}
use  random   values   during   the actual computation to prevent the 
processed data being  exploited by the adversary. Muresan and Gebotys
\cite{muresan04currentflat} proposed a current  flattening technique,
where the dissipated current is flattened  by  adding {\it no-op}s in
the  code  to  provide sufficient discharge. Ambrose et al. \cite{ambrose12randomized} 
proposed a randomized instruction injection technique, where dummy instructions 
were injected during the actual execution. Even though this technique has been proven
effective for small number of data samples, 
the phase  substituition techniques \cite{gebotys07phasesub} can be used
to isolate the injected effects with large number of samples.  
Authors in \cite{Veyrat-Charvillon12Shuffling} present a comprehensive study on
shuffling, which is similar to masking. It was mentioned in \cite{Veyrat-Charvillon12Shuffling}
that the shuffling is effective when both the execution order and the physical resource usage are randomized.

Non-deterministic processing and hardware balancing for power are the
most  explored hardware-based  countermeasures against power analysis
attacks. {\it Non-Deterministic Processors} \cite{May01NDICP} execute
instructions  out-of-order, issuing  independent    code segments
randomly during  runtime and therefore preventing the  adversary from identifying 
the places where specific instructions are executed.  \emph{Dual-Rail} circuits \cite{sokolov05DualRail}  (or
\emph{Dual-Rail Pre-charge}  (DRP)  logic \cite{popp05balancing})  are
designed to consume constant   power regardless of data processed. In
Dual-Rail circuits, each  logic  circuit     is attached to another
similar logic circuit,  complementing  the discharge occurring in the
original logic circuit due to     bit-flips  \cite{saputra03masking}.
Other  similar       hardware balancing techniques are  presented  in \cite{hwang06coprocdesign,popp05balancing,tiri04ASIC,tiri2006WDDL}.
Residue Number Systems (RNS) based randomized hardware approach is presented in \cite{ambrose13DARNS}, randomly 
chooses the moduli sets to obfuscate the binary double and add operation from the power profile. This technique requires 
large data inputs to gain advantage in performance, since RNS circuits cost in hardware. Furthermore, the proposed
technique in  \cite{ambrose13DARNS} is not complete and the authors have not tested the point double and add for ECC. 
Tanimura and Dutt \cite{tanimura10ExCCel} propose an improvement for WDDL \cite{tiri2006WDDL}
using a simulated annealing approach to automatically generate complementary cells to reduce the information leakage from the
power profile. This circuit level balancing ExCCel approach \cite{tanimura10ExCCel} was further improved in \cite{tanimura12HDRL, tanimura12LRCG}.

\emph{MUTE} \cite{ambrose2011multiprocessor}, a  hardware/software information
balancing technique,  is  a    system    level     microarchitectural
countermeasure to protect against power analysis attacks. \emph{MUTE}
uses two processors to execute the same encryption program with   the
actual and complemented data, balancing  information (against power).
\emph{MUTE} requires minimal software instrumentation  compared    to 
the current flattening technique in  \cite{muresan04currentflat}.  It
utilizes  an   identical   second core of the dual core processor and
therefore  requires   minimal hardware (for additional control logic)
compared to the hardware balancing methods \cite{sokolov05DualRail,
hwang06coprocdesign,Hwang06Secure,   popp05balancing, saputra03masking, 
tiri04ASIC,tiri2006WDDL}. However, \emph{MUTE} protects    the system
only against power analysis attacks.

The  dual  core  secure  embedded  processor system presented in this
paper, \SecureD,  makes    use of      techniques  similar  to that of
\emph{IMPRES} \cite{ragel06impres} to   detect code-injection attacks
and to that of \emph{MUTE} \cite{ambrose2011multiprocessor} to protect against
power  analysis attacks. As far as we are aware, there has been no embedded 
processor system, which is immune to  both   code-injection and power
analysis attacks. 

\subsection{Contributions}
\begin{itemize} 
\item A secure dual core embedded system  is proposed to both  detect
      code-injection  attacks   and   protect  against power analysis
      attacks. As far as we are aware, this is the first time an MPSoC solution is proposed to 
      safeguard against both power analysis and code injection, which are two of the major
      security threats in embedded systems. 
\item The processor is capable of detecting bit-flips including those
      causing control flow errors with minimal latency.
\item An interrupt handling mechanism for an ASIP MPSoC is presented. 
\item A  novel switching and synchronizing  mechanism for information
      balancing to counter power analysis is proposed which was not available in \cite{ambrose2011multiprocessor}.               
\item A rapid simulation and synthesis environment is used for a dual core processor. 
\end{itemize}

\subsection{Limitations and Assumptions}
\begin{itemize} 
 \item Our  technique addresses only multiprocessor embedded systems
       with at least two identical (homogeneous) processors. For heterogeneous multiprocessor systems,
       identical functional units for encryption/decryption has to be used in two of the processors. 
 \item We  assume that our system is self contained with    separate
       memories for each of the processors.
 \item Both processors are clocked by a single source.
\end{itemize}

\section{Processor Architecture} \label{arch}
In this section, we give an overview of \SecureD\ architecture.    We
further discuss how the code-injection detection technique similar to
the  one  presented  in \cite{ragel06impres} and  the  power analysis
protection  similar  to  the one presented in \cite{ambrose2011multiprocessor}
are integrated into \SecureD.

\begin{figure}[ht!]
\centering
\includegraphics[width=8.5cm]{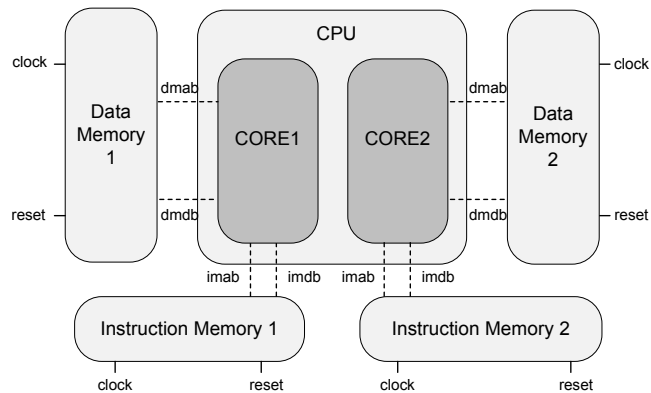}
\caption{The Base (Non Secure) Dual Core Processor with Memory Modules}
\label{baseprocessor}
\end{figure}

Figure~\ref{baseprocessor} depicts  the schematic diagram of the base
dual core processor taken for our design. As depicted, the  processor
has two identical cores with separate instruction and data   memories
for each core. The  base  processor  is statically scheduled to   run
different threads/applications on each of its cores. 

\begin{figure}[hb!]
\centering
\includegraphics[width=8.5cm]{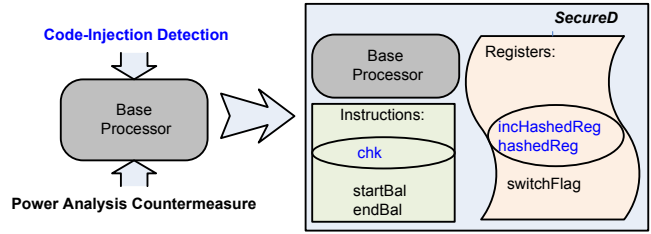}
\caption{Design Alterations to One of the Cores of the Base Processor}
\label{SecureDarch}
\end{figure}

Figure~\ref{SecureDarch}  depicts  how one of the cores from the base
processor  is altered so that it is      capable of detecting  code-%
injection attacks and preventing power analysis attacks. 

As depicted in Figure~\ref{SecureDarch}, each core of our system will
be augmented by adding registers and logic to       handle additional
instructions. The additional  instructions and  registers   encircled
(see Figure~\ref{SecureDarch}) are used      to detect code-injection
attacks and the rest to  prevent   power  analysis attacks. Following
subsections will detail architectural modifications and how they will
be used in ensuring security in our \SecureD\ processor.

\subsection{\SecureD\ for Detecting Code-injection Attacks} \label{impres-arch}
\SecureD\ ensures  code   integrity   of an application by  verifying
whether  all  basic   blocks     of the program are intact at runtime
\cite{ragel06impres}. This is achieved    by         performing basic
block integrity checks at runtime. The basic block  integrity checker
incorporates  the following       tasks: (1) identifying basic blocks
and calculating and assigning checksums  for  each  basic  block   at
compile time; and (2) re-calculating the checksums at runtime     and
comparing them with loaded static values.

At compile time, an application program is grouped into basic  blocks
based on the control flow of the application. Then,  each basic block
is  processed       separately to calculate a checksum  based  on the 
instructions  of  that  block.  The  calculated   checksum    is then
inserted at the beginning  of  each  basic   block  using  a  special
instruction (\emph{chk} instruction). 

At runtime, the first instruction of a loaded   basic     block is  a
\emph{chk}   instruction   that    carries    the    checksum for the 
corresponding  basic  block.  When  an instruction   of  this kind is
fetched, the  checksum   is   loaded    into     a  special  register
(\textit{hashedReg}). The checksum  for each basic block is incrementally
re-calculated  at  runtime while instructions  belonging to the basic
block  are     executed    and  is stored in another special register
(\textit{incHashedReg}). The last instruction of each basic block   is  a
control  flow  instruction  (CFI), and if not, one is inserted at compile time (if  it
is not present at the end of a basic block). CFIs  are  altered  such
that they will (a) incrementally store checksums of their own, and (b)
compare the  result   against          the one loaded through \emph{chk}
instructions. A mismatch  in the comparison  will indicate     a code
integrity violation and generate an exception.

\subsection{\SecureD\ for Preventing Power Analysis Attacks} \label{mute-arch}
When there is an encryption program to be executed, \SecureD\ will use
both of its cores, where the first one will execute      the original
encryption program, while the second core executes  the complementary
program in parallel. Similar to \cite{ambrose2011multiprocessor}, the original
and complementary programs are devised by having the same instruction
sequence  but  complementary  data   processing. Hence, a clock cycle
accurate  synchronization   is     preserved, masking the information
signature in the power profile. The  memories  are  setup statically,
making the processor aware of where the programs are stored and which
data  locations  are  used  for  different  programs. The term
\emph{balancing} is used for this process as was named 
in~\cite{ambrose2011multiprocessor}.

\begin{figure}[ht!]
\centering
\includegraphics[width=7.0cm]{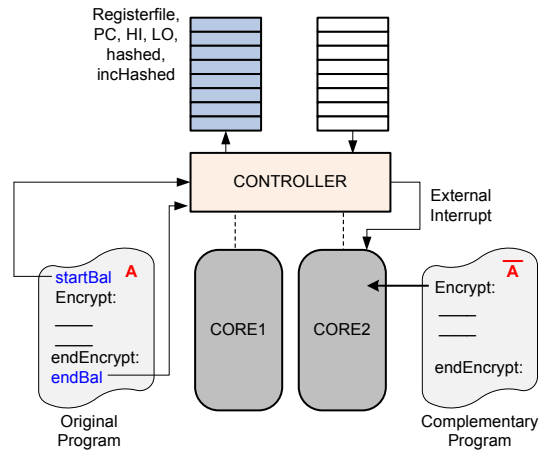}
\caption{Switching and Synchronizing}
\label{balancing}
\end{figure}

The \emph{balancing}  is  triggered     and terminated by two special
instructions which are instrumented in the   source code. As depicted
in             Figure~\ref{balancing},      the  execution  of    the
\emph{startBal}   instruction indicates the \emph{CONTROLLER} that an
encryption program is scheduled in the core. An External Interrupt is 
sent   to      \emph{CORE2}. This is a maskable interrupt which will be
triggered after all the pipelines are flushed. Necessary      registers
(such as the register-file and program counter) of \emph{CORE2}  are saved
in  the  stack     as shown in Figure~\ref{balancing}.  After     the
registers are saved, the \emph{CONTROLLER} sends  the program counter
(PC) values to \emph{CORE1} and \emph{CORE2} on the same clock  cycle
(i.e., PC values of Encrypt in program $A$ and $\overline{A}$).  Both
the original   and complementary programs are executed in parallel by
\emph{CORE1} and \emph{CORE2} respectively. When  the   encryption is
completed, the  \emph{endBal}      instruction in program $A$ will be
executed by \emph{CORE1}, which             will send a signal to the
\emph{CONTROLLER} indicating        the completion of encryption. The
\emph{CONTROLLER}        restores the saved registers from the stack.
\emph{CORE2}  will  resume  its execution and \emph{CORE1}    will be
scheduled with the next program to be executed.

\subsection{Interrupt Handling and \SecureD}
This subsection describes how interrupts are handled by \SecureD\  in
two different scenarios: one, when \SecureD\ is executing  a  regular
application and two, while \SecureD\ is  executing   an    encryption
algorithm (that is during \emph{balancing}). For  both the scenarios,
interrupts will be serviced  after  the pipelines are flushed. Hence,
the registers are updated with correct values. Each interrupt routine
will have three code segments:      one, instructions to save all the
registers into the stack; two, the actual code  for the routine;  and
three, instructions to  restore  the registers from the stack and the
end  instruction to terminate the interrupt. The end        interrupt
instruction    sends a           non-maskable interrupt (NMI)  to the
\emph{CONTROLLER}, which will force the controller to change   the PC
to resume the original execution. In  regular interrupt handling, the
\emph{CONTROLLER} will   change     only the PC of the core which was
servicing  the   interrupt and lets the other core perform its normal
execution.

However, handling  an interrupt          while the core is performing
\emph{balancing} is slightly different. When  a   core    receives an
interrupt during  balancing, the other core   is put on hold until the
interrupt is serviced. The core on hold can also be    allowed to
execute the next  task in the queue, but with careful modification
in the \emph{CONTROLLER}  to  maintain synchronization for balancing.
The NMI request will force the \emph{CONTROLLER} to change the PC  of
both cores to their original location to resume balancing. This is done
on the same clock cycle in-order to  preserve synchronization.

\section{Design Flow} \label{designflow}
In  this  section,  an  overview of the proposed design  flow for the
\SecureD\ architecture is discussed. First, the design of a  software
interface that allows  the      application   to    interact with the
architectural enhancement is described,    and then the design of the
architectural enhancement itself is discussed.

\subsection{Software Design Flow} \label{designflowsw}
\begin{figure}[th!]
\centering
\includegraphics[width=8.5cm]{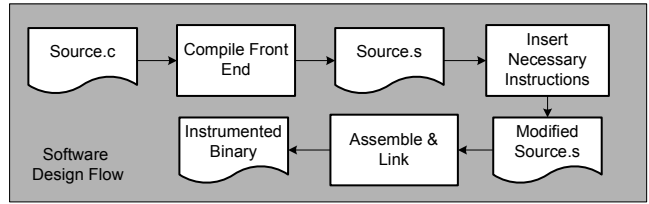}
\caption{Software Design Flow}
\label{swdesignflow}
\end{figure}

Figure~\ref{swdesignflow}     depicts the software design flow of our
framework. The  source  code  of  the   application   program in C is
compiled with the front-end of a compiler to generate   the  assembly
version of the application (\emph{.s files}). Special    instructions
for   both     code-injection detection (\emph{chk} instructions with 
checksum values of basic blocks) and power        analysis protection
(\emph{startBal} and \emph{endBal}) are instrumented into the assembly.
The resulting assembly files are assembled and     linked to generate
the binary (Instrumented Binary in Figure~\ref{swdesignflow}) of  the
application. 

\subsection{Hardware Design Flow} \label{designflowhw}
\begin{figure}[th!]
\centering
\includegraphics[width=8.5cm]{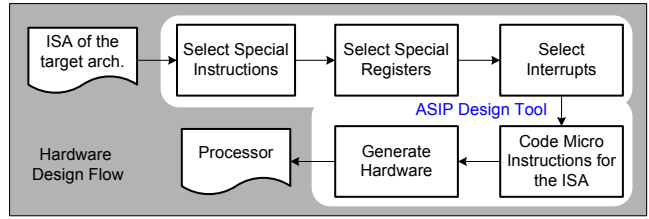}
\caption{Hardware Design Flow}
\label{hwdesignflow}
\end{figure}

Figure~\ref{hwdesignflow}  depicts   the   generation  of a processor
model that implements \SecureD. The  Instruction   Set   Architecture
(ISA)  is  fed    into       an       automatic processor design tool 
(\emph{ASIPMeister~}\cite{ASIPMeister}) to  generate each core of the
processor model. Specifications of special instructions for \SecureD\ 
are given to \emph{ASIPMeister}. Additional registers are chosen   to 
implement  secure    functionalities. Interrupts are selected     and
implemented  for switching and synchronizing  processors. The logical 
functionalities   of      the special instructions and interrupts are
specified as micro-instructions. The single core  processor  model in
VHDL  is       generated using the \emph{generate hardware module} of
\emph{ASIPMeister}. 

\emph{ASIPMeister} generates a synthesizable VHDL description  of   a
single processor model, thus two identical  processors are  generated
separately  and  are   combined    manually to make the VHDL model of
\SecureD. Additional   components   such as the \emph{CONTROLLER}  as
shown in Figure~\ref{balancing} are added during the manual combining
process.

\section{Experimental Setup} \label{experSetup}
\begin{figure}[th!]
\centering
\includegraphics[width=8.5cm]{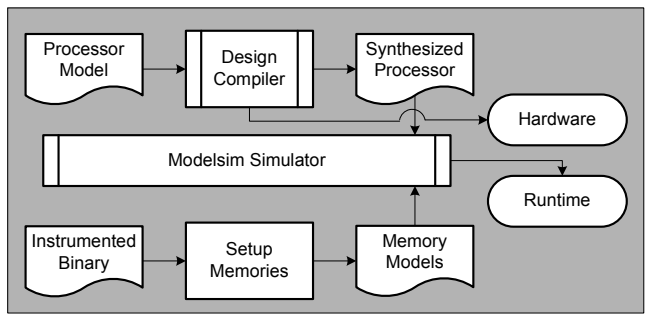}
\caption{Experimental Setup}
\label{expersetup}
\end{figure}

Our \SecureD\ framework is  implemented using  processors with  the PISA
(\emph{Portable Instruction Set Architecture})  instruction   set (as
implemented in \emph{SimpleScalar Tool Set} with  a         six stage
pipeline)            processor without cache. Figure~\ref{expersetup}
illustrates the experimental setup,  identifying     key elements and
tools. Programs in C are compiled using GNU/GCC   cross  compiler for
the PISA instruction set. The \emph{Instrumented  Binary} is produced  as
explained  in     Section~\ref{designflowsw}. The     processor model
represents the \SecureD\ processor in VHDL description. 

Synopsys\reged \emph{Design  Compiler} is     used to  synthesize the
processor model as shown in Figure~\ref{expersetup}. Hardware details
are reported by the \emph{Design Compiler}. The instrumented   binary
is processed to set up  both   data  and    program memories. Modelsim
VHDL simulator is used to simulate the synthesized processor  and the
memory models to deduce the runtime (in number of clock cycles).

\section{Results} \label{results}
This    section presents the hardware and runtime overhead details of
our \SecureD\ processor,   comparing  with the processors implemented
for only  detecting code-injection attacks - \emph{SecureD\nobreakdash-I}
\cite{ragel06impres}, and for only protecting  against  power analysis
attacks - \emph{SecureD\nobreakdash-M} \cite{ambrose2011multiprocessor}.

\subsection{Hardware Summary} \label{hwsumm}
Table~\ref{hwdetails}   tabulates    the hardware summary produced by
Design Compiler  for       all      processors investigated. The CMOS 65nm technology is used.
\emph{Non-SecureD} is a processor with dual cores and   without   any
secure implementations. \emph{SecureD\nobreakdash-M}    costs        an
additional  1.9\%    and 2.7\% in gates and cells respectively, while
\emph{SecureD\nobreakdash-I}   costs   an     additional 1.3\% and 1.5\%
compared  to   the   \emph{Non-Secure} processor. \SecureD\ 
costs 3.8\% area overhead in both gates and cells.

\begin{table}[ht!]
\centering
\caption{Hardware Details} \label{hwdetails}
\begin{tabular}[t]{|l|r|r|r|}\hline
                  &{\bf Area}     &{\bf Cell Area}&{\bf Clock}\\
                  &{(\# of gates)}&{(\# of cells)}& {(ns)} \\ \hline\hline
\emph{Non-SecureD}& 484180        & 132828        & 31.58  \\ \hline
\emph{SecureD-M}  & 493330        & 136496        & 31.06  \\ \hline
\emph{SecureD-I}  & 490715        & 134812        & 31.94  \\ \hline
\emph{SecureD}    & 502830        & 137906        & 31.80  \\ \hline
\end{tabular}
\end{table}

The \emph{SecureD\nobreakdash-M} processor has a slight decrease in the clock width of  the
critical path because of the optimizations    performed by the Design
Compiler. \emph{SecureD\nobreakdash-I} has a slight increase in the clock due to its
runtime checksumming logic. \SecureD\     expectedly has an
increased clock width compared to \emph{Non-SecureD} and   \emph{SecureD\nobreakdash-M}
  and    a           slightly decreased clock width compared to
\emph{SecureD\nobreakdash-I} because of the optimizations.  

\subsubsection*{Hardware Summary in FPGA} \label{fpgaimpl}
The base dual   core  and \SecureD\ processors were implemented on an
FPGA (XC2V3000-4FG676) and the  hardware details       are
presented  in Table~\ref{hwsumFPGA}. \SecureD\ consumes an additional
10.8\% in       slices, 2.8\% in LUTS, 1.5\% in IOBs compared  to the
non-secure dual    core processor \emph{Non-SecureD}. Clock width has
slightly increased in addition to the hardware overhead in logics, such as LUTs, IOBs and slices.  

\begin{table}[ht!]
\centering
\caption{Hardware Details in XC2V3000-4FG676} \label{hwsumFPGA}
\begin{tabular}[t]{|l|r|r|}\hline
                          &{\bf \emph{Non-SecureD}}&{\bf \SecureD}\\ \hline\hline
         {\bf Hardware}   &                        &        \\ 
~~~Slice Flip-flops (28672)& 4300                  &   4823 \\ 
~~~4 input LUTs (28672)   & 16476                  &  16942 \\ 
~~~Bonded IOBs (842)      &   410                  &    416 \\ 
~~~GCLKs (16)             &     1                  &      1 \\ \hline
           {\bf Timing}   &                        &        \\
~~~Clock (ns)             &   209                  &    216 \\
~~~Logic Levels           &   515                  &    529 \\ \hline\hline
\end{tabular}
\end{table}

\subsection{Runtime Analysis} \label{runtime}
\begin{figure*}[ht!]
\centering
\includegraphics[width=12.2cm]{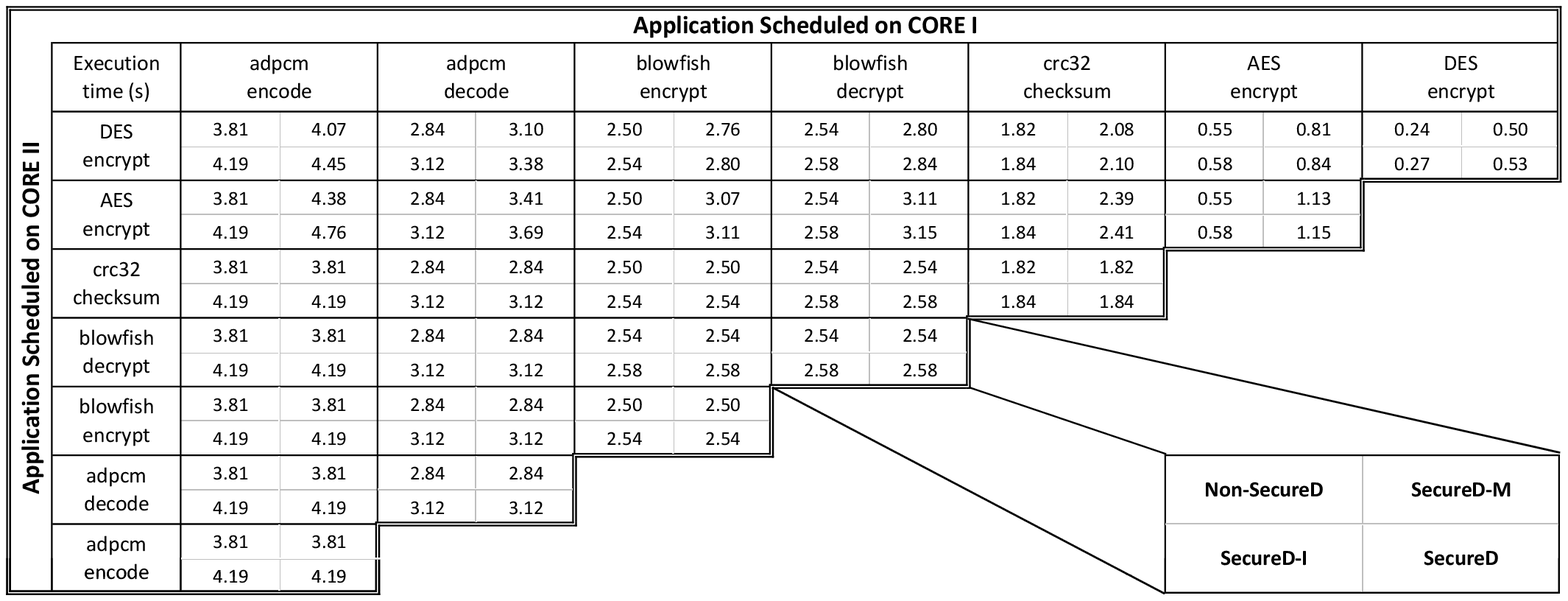}
\vspace{1mm}
\caption{Runtime Comparisons}
\label{runtimeanal}
\end{figure*}

Figure~\ref{runtimeanal}   depicts   the runtime analysis of all four
configurations of our dual  core      processors (\emph{Non-SecureD},
\emph{SecureD\nobreakdash-M}, \emph{SecureD\nobreakdash-I} and \SecureD)     for   different
benchmark applications. Out of the seven applications tabulated,  two
(DES and AES) are  considered to be encryption algorithms and 
\emph{balancing} is enabled while they are executed. Each cell in the
table (see Figure~\ref{runtimeanal}) has 4 parts, where  the top left part
specifies the runtime of a particular  application  on the non-secure
dual core (\emph{Non-SecureD}), the top right part represents  the runtime
on \emph{SecureD\nobreakdash-M} processor, the bottom left     part indicates the
runtime on \emph{Secured\nobreakdash-I} processor and  the bottom right part  has
the runtime of an application on \SecureD\  processor. The \emph{Non-SecureD} 
neither includes the balancing feature nor the code injection detection feature,
whereas the \emph{SecureD\nobreakdash-I} only includes the code injection detection feature.
The \emph{SecureD\nobreakdash-M} design only contains the balancing solution to prevent power analysis
whereas the \SecureD\ processor prevents both the attacks. 

When two applications are scheduled on dual cores, the runtime of the
application which is higher is considered for the net runtime      of
the  two  applications.  Hence,  the  cells along the diagonal of the
table (in Figure~\ref{runtimeanal}) represent   the  runtime  of each
application alone.           Since  \emph{adpcm.encode}       has the
highest runtime amongst all            programs,    combinations with
\emph{adpcm.encode} will be having the runtime of \emph{adpcm.encode}
(e.g., see \emph{Non-SecureD} cells  in               column two   of 
Figure~\ref{runtimeanal}). 

In \emph{SecureD\nobreakdash-M} and \SecureD\ configurations, when either AES  or
DES (encryption algorithms) is scheduled to one of the cores,     the
application scheduled   on      the second core has to wait until the 
\emph{balancing} of  AES   or         DES comes to an end (i.e., when
\emph{balancing} is performed, both cores  will be executing the same
program, either AES or DES, with complementary values).   Note  that,
currently, \emph{balancing} is performed only for AES and DES in  the
given  set     of benchmark applications. For example, the runtime of 
applications \emph{adpcm.encode} and \emph{DES encrypt} scheduled  on
\emph{SecureD\nobreakdash-M} (see Figure~\ref{runtimeanal} row 2, column 2),   is
the    addition    of the runtime of \emph{DES encrypt} on both cores
(for \emph{balancing}), the time taken for switching, and the runtime
of \emph{adpcm.encode} on one   of the cores which is performed after
the \emph{balancing} finishes.

As depicted in Figure~\ref{runtimeanal}, the cells corresponding   to
\emph{SecureD\nobreakdash-I} consume additional time due to the runtime   check-%
summing used for detecting code integrity violations.     The average
runtime  overhead  of \emph{SecureD\nobreakdash-I} for all  seven applications is
6.06\%. Compared to the runtime of \emph{Non-SecureD}  configuration,
\emph{SecureD\nobreakdash-M} consumes extra runtime only when either AES or   DES
is  scheduled on one of or both the cores. The   average      runtime
overhead of all seven applications on \emph{SecureD\nobreakdash-M} is      15.4\%
compared to \emph{Non-SecureD}. As \SecureD\ integrates          both 
code-injection detection and power analysis protection, it   consumes
extra runtime similar to that of \emph{SecureD\nobreakdash-I} when no application
with  encryption (such as AES or DES) is present. \SecureD\  consumes
additional  runtime compared to \emph{SecureD\nobreakdash-I} when    applications 
such as AES or DES are scheduled to one or both of the     cores. The
average runtime    overhead of \SecureD\ with two encryption and five
regular applications is 20.0\%.

Table~\ref{intrdelay} presents    the delay in number of clock cycles
for   switching   and    interrupt servicing during balancing. All 37
registers including the 32   registers in the register-file are saved
and restored sequentially. Every time  an external interrupt is fired
(to switch to balancing or to service interrupt) there  is  a 6 clock
cycle  delay  to  flush  the pipeline to update all the  registers. A
clock  cycle  delay  is    consumed for each interrupt call or switch
instruction (\emph{startBal}). Another clock cycle delay is needed to
exit the interrupt or to   exit \emph{balancing} (using \emph{endBal}
instruction).

\begin{table}[ht!]
\centering
\caption{Switching and Interrupt Servicing Delay} \label{intrdelay}
\begin{tabular}[t]{|l|r|}\hline
           & \multicolumn{1}{|c|}{\bf Clock Cycles} \\\hline\hline
{\bf Delay}                    &      \\
~~~Store Register-file         &  320 \\
~~~Store PC,HI,LO registers    &   30 \\
~~~Store incHashed, hashed     &   20 \\
~~~Restore Register-file       &  320 \\
~~~Restore PC,HI,LO registers  &   30 \\
~~~Restore incHashed, hashed   &   20 \\
~~~Flush Pipelines             &    6 \\
~~~Interrupt to switch         &    1 \\
~~~Exit the interrupt          &    1 \\\hline
{\bf Total Delay}              &  748 \\\hline\hline
\end{tabular}
\end{table}

\section{Discussions} \label{discussions}
This paper focuses on power analysis based side channel attacks. Our solution can be further applied 
to electromagnetic based side channel attacks \cite{homma10electro}, however not tested. 
Previous techniques have proven that the DPA resistant techniques have also prevented DPA, since
electro-magnetic emmissions are closely related to power dissipations \cite{gebotys06masking, gebotys07phasesub}.
 
We assume that the processors are designed and layed out as homogeneous components which will dissipate 
around the same amount of power. However, this could be a challenging task and would require manual 
design considerations. We could use the CoRaS solution \cite{ambrose12CoRaS} to counteract this limitation
by randomly swapping rounds of the encryption/decryption between processors. 

We assume that we do not have caches in our MPSoC or disable caches during the balancing operation. This is to make
sure we execute both the processors in instruction-level lock-step mode. Cache locking mechanisms \cite{asaduzzaman10improving} can be utilized 
during secure execution, if we want to implement cache to enhance performance.

\section{Conclusions} \label{conclusions}
In  this  paper,  we  have  presented \SecureD, a dual core     based
multiprocessor, which is immune to code-injection attacks and   power
analysis   based     side-channel     attacks. Both these attacks are
considered to be the most successful attacks in embedded systems   in
recent years. \SecureD\ protects against code-injection by performing
static time basic block instrumentation (with checksums)  and runtime
comparison for detecting code  integrity violations        and  power
analysis  is prevented by implementing information balancing  to mask
the actual information (or the secret key) from the power    profile.
\SecureD\ is easily scalable beyond dual processors and has   minimal
performance and hardware  overhead    compared to previous techniques
proposed to separately handle the attacks.

\bibliographystyle{abbrv}
\bibliography{referenence}     
\end{document}